\documentclass[twocolumn,showpacs,preprintnumbers,amsmath,amssymb]{revtex4}
\usepackage{amsmath,amssymb,graphics,epsfig,subfigure}
\usepackage{color}

\begin{document}

\thispagestyle{empty}

\begin{center}

\title{Repulsive Interactions and Universal Properties of Charged Anti-de Sitter Black Hole Microstructures}

\date{\today}
\author{Shao-Wen Wei$^{1,2}$ \footnote{E-mail: weishw@lzu.edu.cn} ,
        Yu-Xiao Liu$^{1}$ \footnote{E-mail: liuyx@lzu.edu.cn} , and Robert B. Mann$^{2}$ \footnote{E-mail: rbmann@uwaterloo.ca}}

 \affiliation{ $^{1}$Institute of Theoretical Physics $\&$ Research Center of Gravitation, Lanzhou University, Lanzhou 730000, People's Republic of China\\
$^{2}$Department of Physics and Astronomy, University of Waterloo, Waterloo, Ontario, Canada, N2L 3G1}

\begin{abstract}
The Ruppeiner geometry of thermodynamic fluctuations provides a powerful diagnostic of black hole microstructures. We investigate this for charged AdS black holes and find that while an attractive microstructure interaction dominates for most parameter ranges, a weak repulsive interaction dominates for small black holes of high temperature. This unique property distinguishes the black hole system from that of a Van der Waals fluid, where only attractive microstructure interactions are found. We also find two other novel universal  properties for charged black holes. One is that the repulsive interaction is independent of the black hole charge and temperature. The other is that the behavior of the Ruppeiner curvature scalar near criticality is characterized by a dimensionless constant that is identical to that for a Van der Waals fluid, providing us with new insight into  black hole microstructures.
\end{abstract}


\maketitle
\end{center}

{\it Introduction}
Understanding the microscopic structure of black holes has been an important subject since the establishment of the four laws of black hole thermodynamics \cite{Hawking,Bekensteina,Bekensteinb,Bardeen}, with surface gravity and area respectively identified as black hole temperature and entropy. The latter differs from an ordinary thermodynamic system, in that it is proportional to an area (of the event horizon) rather than a volume. Many different kinds of approaches and theories of gravity have been explored with an aim of understanding this interesting property \cite{Vafa,Maldacena,Callan,Horowitz,Emparan,Lunin,Mathur}. The salient point is to discern the underlying degrees of freedom of a black hole, i.e. its microscopic structure. A black hole of nonvanishing Hawking temperature must possess its own microscopic structure if one accepts Boltzmann's insight: if you can heat it, it has microscopic structure.

For an ordinary fluid system one can start from its microscopic molecular constituents, and then construct its macroscopic thermodynamic quantities following statistical mechanics. However given our present state of knowledge,  the inverse process -- a thermodynamic geometric approach -- must be carried out  to discern the microscopic structure of a black hole.  We shall start with one of the mainstays of statistical mechanics: the entropy formula given by Boltzmann
\begin{equation}
 S=k_{\rm B}\ln\Omega,
\end{equation}
where the Boltzmann constant $k_{\rm B}\approx 1.38\times 10^{-23} J/K$, and $\Omega$ is the number of the microscopic states of the corresponding thermodynamic system. Inverting this yields
\begin{equation}
 \Omega=e^{\frac{S}{k_{\rm B}}},\label{Os}
\end{equation}
which is   the starting point of thermodynamic fluctuation theory. For a system of $N+1$ independent variables $x^{\mu}$ with $\mu$=0, 1, ..., $N$, the probability of finding its state to be between $(x^{0}, ..., x^{N})$  and $( x^{0}+ dx^{0}, ..., x^{\rm N} + dx^{\rm N})$ is proportional to the number of microstates
\begin{equation}
 P(x^{0}, ..., x^{\rm N})dx^{0}\cdot \cdot \cdot dx^{\rm N}=
 C\Omega (x^{0}, ..., x^{\rm N})dx^{0}\cdot \cdot \cdot dx^{\rm N},
\end{equation}
where $C$ is a normalization constant. Hence
\begin{equation}
  P(x^{0}, ..., x^{\rm N})  \propto e^{\frac{S}{k_{\rm B}}}.
  \label{pp}
\end{equation}
Consider a thermodynamic system partitioned into a small sub-system S plus its remainder regarded as the environment E. The total entropy can be written as $S(x^{0}, ..., x^{\rm N})=S_{\rm S}(x^{0}, ..., x^{\rm N})+S_{\rm E}(x^{0}, ..., x^{\rm N})$ with $S_{\rm S}\ll S_{\rm E}\sim S$. Expanding the total entropy near its local maximum at $x^{\mu} = x^{\mu}_0$, we have
\begin{align}
&S=S_{0} + \left. \frac{\partial S_{\rm S}}{\partial x^{\mu}}  \right|_{0} \Delta x^{\mu}_{\rm S}
+ \left.\frac{\partial S_{\rm E}}{\partial x^{\mu}}   \right|_{0}   \Delta x^{\mu}_{\rm E}
\label{qwq}\\
&+ \left. \frac{1}{2}\frac{\partial^{2}S_{\rm S}}{\partial x^{\mu}\partial x^{\nu}}
        \right|_{0}  \Delta x^{\mu}_{\rm S}\Delta x^{\nu}_{\rm S}
       + \left. \frac{1}{2}\frac{\partial^{2}S_{\rm E}}{\partial x^{\mu}\partial x^{\nu}}
        \right|_{0}  \Delta x^{\mu}_{\rm E}\Delta x^{\nu}_{\rm E}
   +\cdots \nonumber
\end{align}
where $S_{0}$ measures the local maximum of the entropy and ``$|_{0}$'' means $|_{x^{\mu} = x^{\mu}_0}$. It is natural to suppose that the fluctuating parameters are conservative and additive i.e., $x^{\mu}_{\rm S}+x^{\mu}_{\rm E}=x^{\mu}_{\rm total}=$constant; hence $\frac{\partial S_{\rm S}}{\partial x^{\mu}} |_{0} \Delta x^{\mu}_{\rm S}=-\frac{\partial S_{\rm E}}{\partial x^{\mu}} |_{0} \Delta x^{\mu}_{\rm E}$. Noting  that $S_{\rm E}$ is of the same scale as that of the total system, the last term of (\ref{qwq}) is much smaller than the fourth term, and one can ignore it. Hence
\begin{eqnarray}
 \Delta S=S-S_{0} = \left. \frac{1}{2}\frac{\partial^{2}S_{\rm S}}{\partial x^{\mu}\partial x^{\nu}}
        \right|_{0}  \Delta x^{\mu}_{\rm S}\Delta x^{\nu}_{\rm S}
        +\cdots.
\end{eqnarray}
Absorbing $S_{0}$ into the normalization constant, we have the probability
\begin{equation}
 P(x^{0}, ..., x^{\rm N}) \propto e^{-\frac{1}{2}\Delta l^{2}},
\end{equation}
where
\begin{eqnarray}
 \Delta l^{2}&=&-\frac{1}{k_{\rm B}}\frac{\partial^{2}S_{\rm S}}{\partial x^{\mu}\partial x^{\nu}} \Delta x^{\mu}\Delta x^{\nu}
 \label{Ds}
\end{eqnarray}
is the distance  between two neighboring fluctuation states \cite{Ruppeiner}.

From the perspective of thermodynamic information geometry, the less probable a fluctuation between two thermodynamic states, the further apart they are. Thus this line element \eqref{Ds} encodes information about the effective interaction between two microscopic fluctuation states. For a given fluid system, the scalar curvature of its information metric \eqref{Ds} is an indicator of its
microstructure interactions \cite{Ruppeiner,Oshima}: positive/negative scalar curvature respectively implies that a repulsive/attractive interaction dominates, whereas vanishing curvature indicates   repulsive and attractive interactions are in balance. Furthermore, it is natural to conjecture that the value of the scalar curvature measures the strength of the interactions.  The thermodynamic potential is clearly the entropy of this information geometry, which is known as the Ruppeiner geometry.

{\it Ruppeiner geometry and microstructures}
Henceforth we set $k_{\rm B}=1$ and drop the index S in the entropy $S_{\text{S}}$. Our aim is to employ the tools of Ruppeiner geometry to the charged AdS black hole system to probe its microstructure interactions, taking the temperature $T$ and thermodynamic volume $V$ as the fluctuation variables. 

\begin{figure}
\includegraphics[width=6cm]{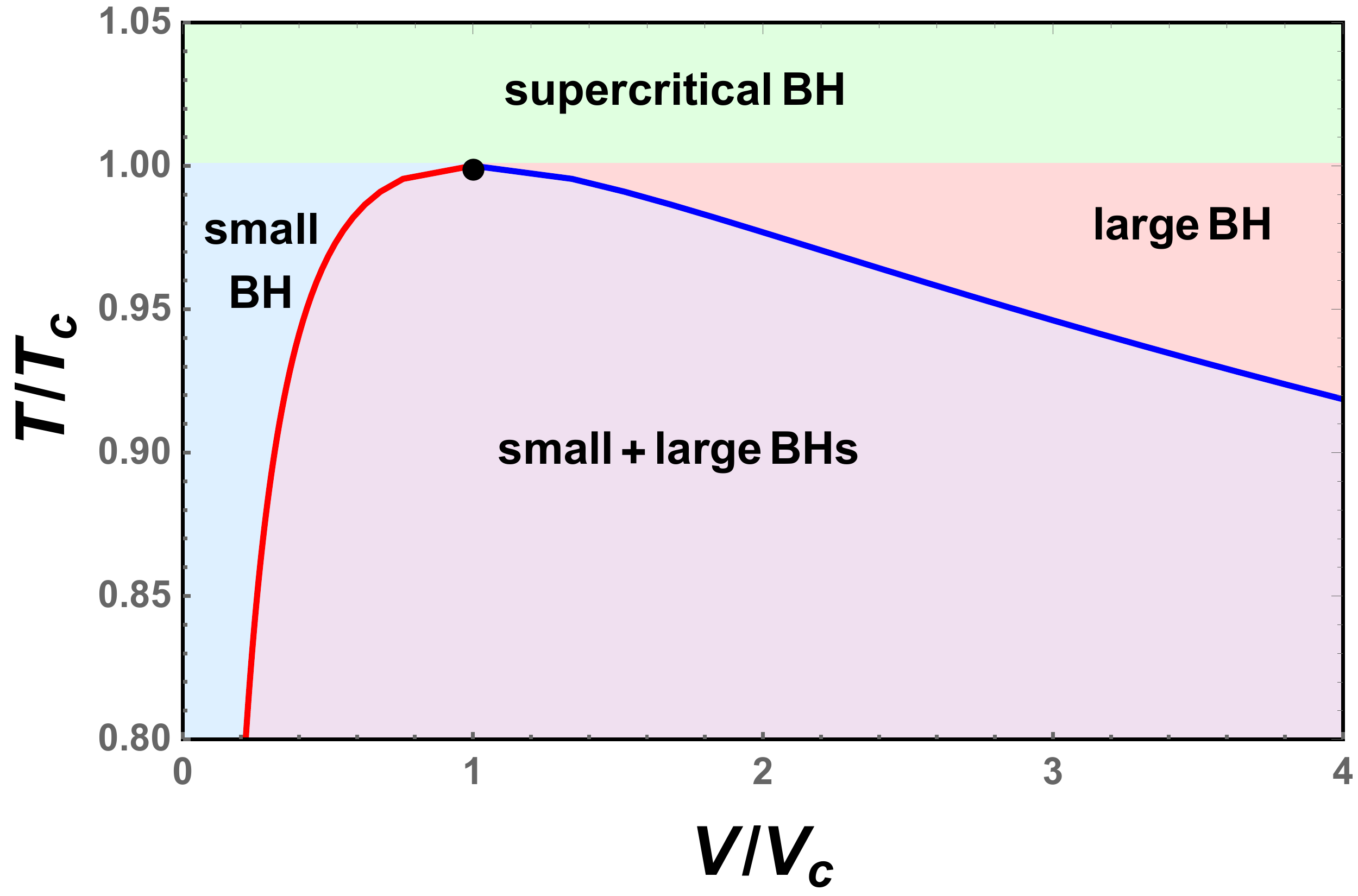}
\caption{Phase structure of the charged AdS black hole in $\tilde{T}$-$\tilde{V}$ diagram. Red (left) and blue (right) solid curves separated by the critical point (black dot) correspond to saturated small and large black holes. Note that in the coexistence region (light purple) of small and large black holes, the equation of state is not applicable.}\label{ppss}
\end{figure}

First, we express the line element (\ref{Ds}) as
\begin{equation}
 dl^{2}=\frac{C_{V}}{T^{2}}dT^{2}+\frac{(\partial_{V}P)_{T}}{T}dV^{2}
 \label{xxy}
\end{equation}
using  the  thermodynamic first law for charged AdS black holes, where
$C_{V}=T\big(\frac{\partial S}{\partial T}\big)_{V}$ is the heat capacity at constant volume.
The  equation of state reads \cite{Kubiznak}
\begin{equation}
 P=\frac{T}{v}-\frac{1}{2\pi v^{2}}+\frac{2Q^{2}}{\pi v^{4}}, \label{stateeq}
\end{equation}
where the specific volume $v=2r_{\rm h}$, with $r_{\rm h}$ the horizon radius. This system is known to exhibit a small-large black hole phase transition, where the charge $Q$  governs the critical point: $ P_{\rm c}=1/96\pi Q^{2}$, $T_{\rm c}=\sqrt{6}/18\pi Q$, $v_{\rm c}=2\sqrt{6}Q$. The small/large black hole coexistence curve has the analytic form \cite{Spallucci}
\begin{eqnarray}
 \tilde{T}^{2}=\tilde{P}(3-\sqrt{\tilde{P}})/2,\label{tt}
\end{eqnarray}
where $\tilde{T}=T/T_{\rm c}$ and $\tilde{P}=P/P_{\rm c}$ are the respective reduced temperature and pressure.  From this we can construct a phase diagram, shown in Fig. \ref{ppss}, in the $\tilde{T}$-$\tilde{V}$ plane, {where $\tilde{V}=V/V_{\rm c}$ with $V=\frac{4}{3}\pi r_{\rm h}^{3}$ and $V_{\rm c}=8\sqrt{6}\pi Q^{3}$}.
Small and large black hole phases are respectively located at the left and right. The supercritical black hole phase is above the critical point marked with a black dot. The red (left) and blue (right) solid curves respectively describe saturated small and large black holes, and below these curves is the coexistence phase. One key feature of the coexistence region is that the equation of state (\ref{stateeq}) does not apply.

One special property of charged AdS black holes is that their heat capacity at constant volume vanishes, i.e., $C_{V}$=0. This renders the line element (\ref{xxy}) singular, and consequently information of the associated black hole microstructure is not revealed from the thermodynamic geometry. We shall deal with this by treating $C_{V}$ as a constant whose value is infinitesimally close to zero, defining a new normalized scalar curvature $R_{\rm N}$
\begin{equation}
 R_{\rm N}=R C_{V}
\end{equation}
from the Ruppeiner curvature scalar $R$.  Noting that the heat capacity of a Van der Waals (VdW) fluid $C_{v}=\frac{3}{2}k_{\rm B}$ is of order $10^{-23}$, in effect we are treating the black hole's vanishing heat capacity as a $k_{\rm B}\rightarrow0^{+}$ limit. In what follows we show that $R_{\rm N}$ provides useful information regarding properties of the  black hole's microstructure.

\begin{figure}
\includegraphics[width=6cm]{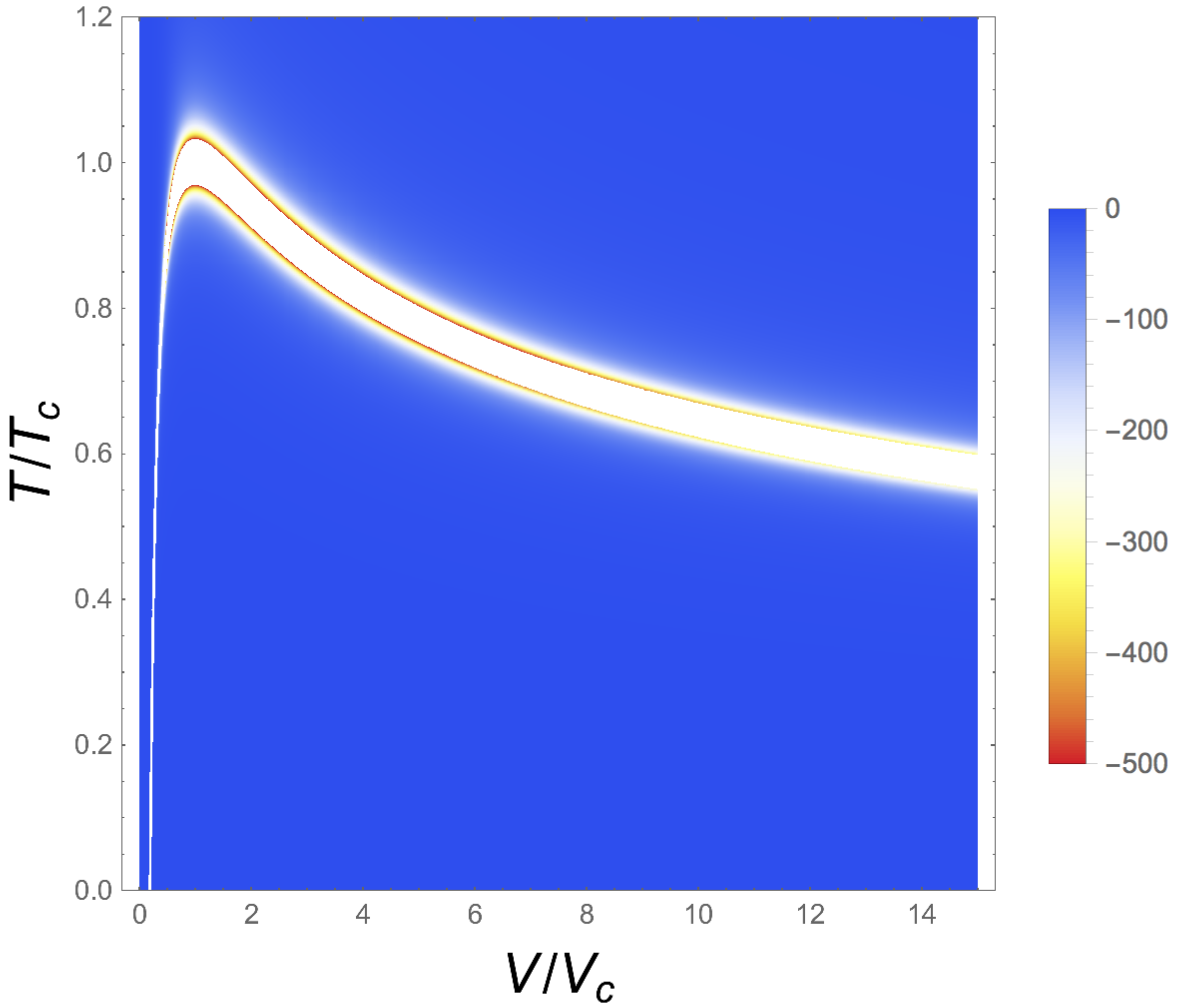}
\caption{Behavior of the normalized scalar curvature $R_{\rm N}$.}\label{pRuppC}
\end{figure}

After a simple calculation, we find
\begin{equation}
 R_{\rm N}=\frac{(3\tilde{V}^{\frac{2}{3}}-1)(3\tilde{V}^{\frac{2}{3}}-4\tilde{T}\tilde{V}-1)}{2(3\tilde{V}^{\frac{2}{3}}-2\tilde{T}\tilde{V}-1)^{2}}
 \label{crnn}
\end{equation}
for the normalized scalar curvature $R_{\rm N}$. {Note that  $R_{\rm N}$ does not explicitly} depend on the black hole charge $Q$--all black holes with different charge share the same expression in the reduced parameter space, a universal result.  We depict the behavior of $R_{\rm N}$   in Fig. \ref{pRuppC}. We see that for most of the parameter space the value of $R_{\rm N}$ is near zero.

However  near the curve
\begin{equation}
 \tilde{T}_{\rm div}=\frac{3\tilde{V}^{\frac{2}{3}}-1}{2\tilde{V}},\label{spcurve}
\end{equation}
$R_{\rm N}$ changes dramatically, and on this curve it goes to negative infinity. This behavior also implies that the black hole microstructure changes quickly in the vicinity of the temperature $\tilde{T}_{\rm div}$. We also compute the curves where $R_{\rm N}$ changes sign
\begin{eqnarray}
 \tilde{T}_{0}&=&\frac{3\tilde{V}^{\frac{2}{3}}-1}{4\tilde{V}},\label{sc1}\\
 \tilde{V}_{0}&=&\frac{1}{3\sqrt{3}},\label{sc2}
\end{eqnarray}
whose traversal indicates a change between attractive or repulsive interactions of the microstructure. These curves are universal, applying to all charged AdS black holes  regardless of the value of $Q$.

To better understand the microstructure interactions we compare the situation to  the VdW fluid system. The interaction between two neighboring VdW fluid molecules  is given  by the Lennard-Jones potential, which describes a short-range repulsive  and longer-range attractive interaction. In the ``hard-core" model, the size of the molecules is generally chosen to be equal to the equilibrium point of the interaction \cite{Johnston}, thereby excluding the short-range repulsive interaction. After taking the ``mean-field" approximation the attractive part  dominates in the fluid system. Nevertheless repulsive interactions could exist due to thermal effects or molecular collisions.

\begin{figure}
\includegraphics[width=6cm]{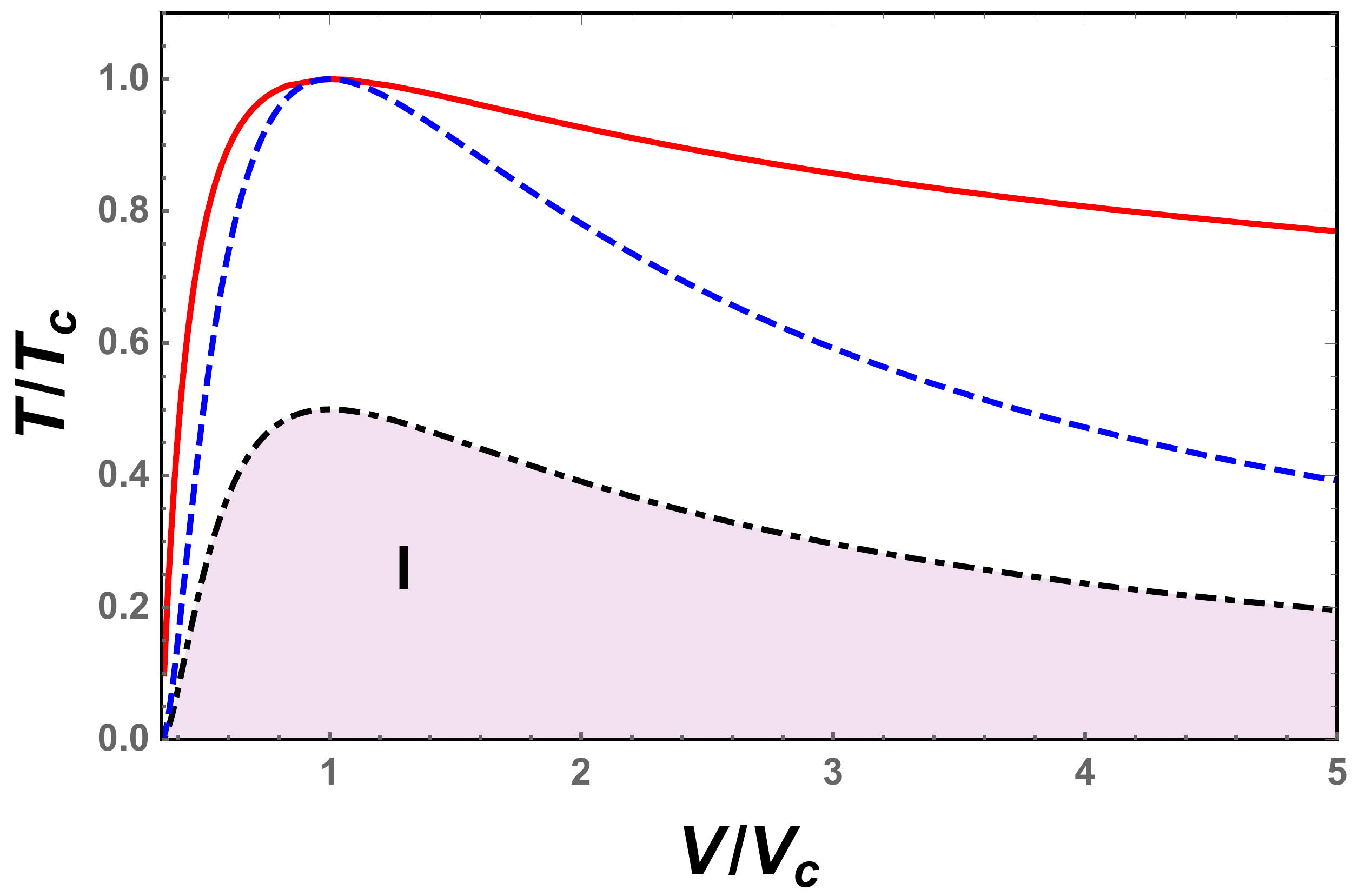}
\caption{Characteristic curves for the VdW fluid. The coexistence and  sign-changing curves are respectively described by the red solid and black dotted-dashed lines. The blue dashed line corresponds to the temperature $\tilde{T}_{\rm div}$, on which $R \to -\infty$. In the shaded region I, $R > 0$;  otherwise,  $R<0$.}\label{pVdWtv}
\end{figure}

In Fig. \ref{pVdWtv} we illustrate the coexistence (red solid line) and sign-changing (black dotted-dashed line) curves of the VdW liquid and gas phases, as well as the curve $\tilde{T}_{\rm div}$ (blue dashed line) at which $R$ diverges. Region I (shaded, below the sign-changing curve) has positive scalar curvature $R$ indicating that the repulsive interaction dominates in this region. Otherwise, $R$ is negative and the attractive interaction dominates. Furthermore, as shown in Fig. \ref{pRuppC}, significant changes in $R$ occur only near $\tilde{T}_{\rm div}$, and so a weak attractive interaction dominates for the VdW fluid system far away from the critical case. However we emphasize that the
equation of state is inapplicable below the coexistence line, and so any features appearing in the coexistence region (such as region I) are tentative. We can state with confidence that an attractive microstructure interaction is always dominant for a VdW fluid above the coexistence line.


\begin{figure}
\includegraphics[width=6cm]{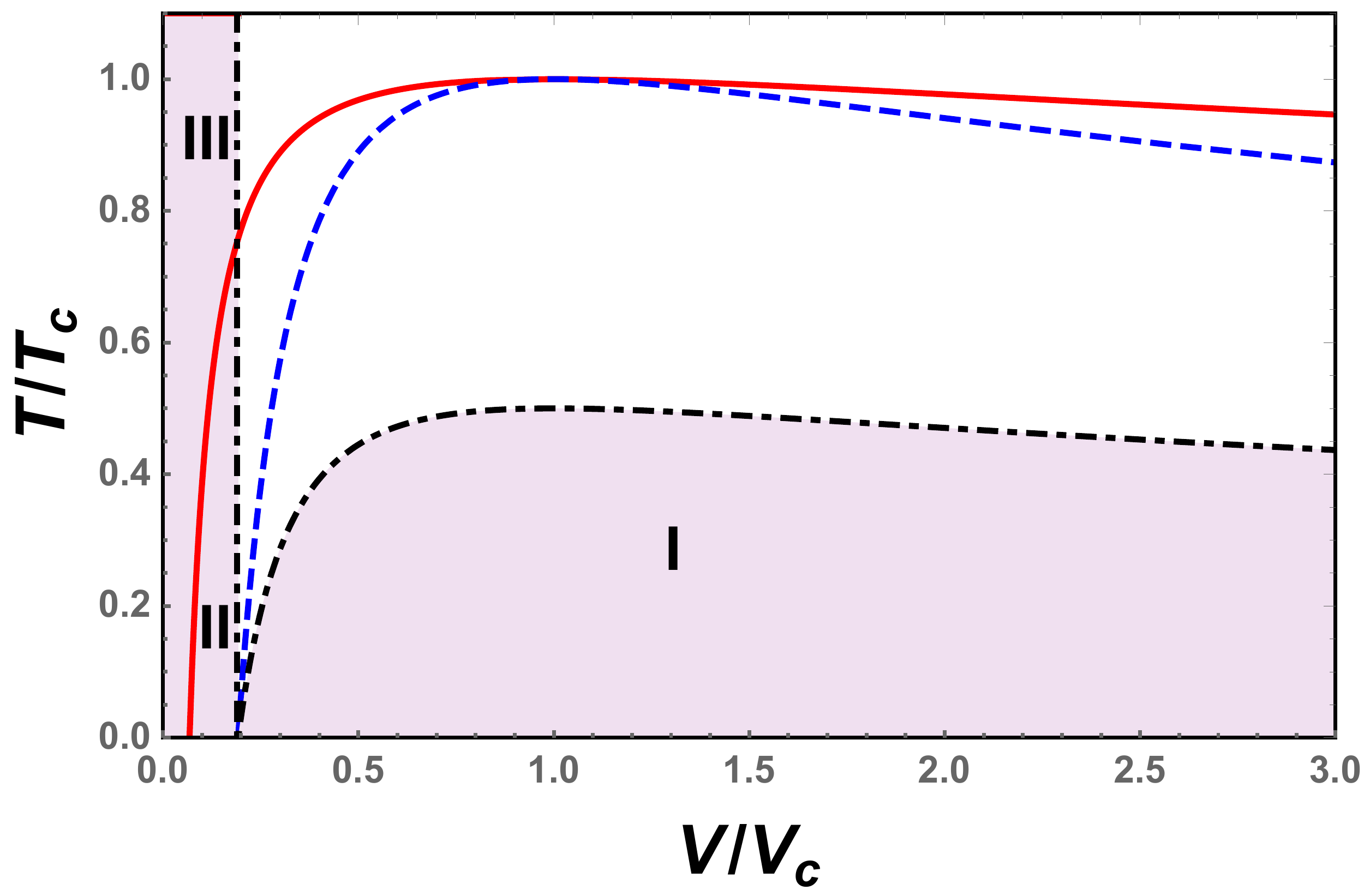}
\caption{Characteristic curves for the charged AdS black hole. These have the same meaning as those in Fig. \ref{pVdWtv}. In the shaded regions I, II, and III, the scalar curvature $R_{\rm N}> 0$;  otherwise $R_{\rm N}<0$.}\label{pCAdStv}
\end{figure}

In Fig. \ref{pCAdStv} we depict the analogous   diagram for the charged AdS black hole, whose phase transition behavior has been known for some time to be qualitatively the same as that of a VdW fluid \cite{Kubiznak}.  Using (\ref{sc1}) and (\ref{sc2})  the coexistence curve for small and large black hole phases can be plotted, along with the sign-changing and $\tilde{T}_{\rm div}$ curves for the charged AdS black hole, all shown in Fig. \ref{pCAdStv} with the same format. It is easy to see that the phase structure is  quite similar to the VdW system at large $\tilde{V}$; however for small $\tilde{V}$ the situation is markedly different.  In addition to region I, two more regions II and III also have positive $R_{\rm N}$.  As with the VdW fluid, the equation of state is inapplicable below the coexistence curve, rendering the existence of  regions I and II tentative. However the existence of region III, above the coexistence curve is robust. Consequently repulsive interactions dominate among the microstructures of charged AdS black holes of small volume at sufficiently high temperature, in strong contrast to the situation for a VdW fluid. Since region III is far from the $\tilde{T}_{\rm div}$ curve, $R_{\rm N}$ is small and so the repulsive microstructure interaction, while dominant, is weak. A weak attractive interaction dominates in other parameter regions above the coexistence curve.

 {\it Critical behavior of the normalized scalar curvature}
As shown above, despite the inapplicability of the equation of state in the coexistence region, $R_{\rm N}$ can diverge at the critical point. We can expand the normalized scalar curvature $R_{\rm N}$ along the saturated small and large black hole curves near the critical point, obtaining
\begin{eqnarray}
 R_{\rm N}({\rm SBH})&=&-\frac{1}{8}t^{-2}+\frac{1}{2\sqrt{2}}t^{-\frac{3}{2}}+\mathcal{O}(t^{-1}),\label{rnsb1}\\
 R_{\rm N}({\rm LBH})&=&-\frac{1}{8}t^{-2}-\frac{1}{2\sqrt{2}}t^{-\frac{3}{2}}+\mathcal{O}(t^{-1}),\label{rnsb2}
\end{eqnarray}
for the small (SBH) and large (LBH) black hole cases respectively, where $t=1-\tilde{T}$. We see that $R_{\rm N} \to -\infty$ at the critical point with a universal critical exponent of $2$. Noting that near the critical point, the correlation length $\xi\sim t^{-\nu}$, we conclude
\begin{equation}
R_{\rm N}  \sim-\xi^{\frac{2}{\nu}}   \sim   -\xi^{4}
\end{equation}
where the latter relation follows from mean field theory, for which $\nu=1/2$. Since other critical phenomena, such as critical opalescence, are closely linked with $\xi$, we expect this phenomenon could have a geometric interpretation.
Further investigation should provide us with novel insight into these different critical phenomena from a thermodynamic geometric perspective.

Moreover, from Eqs. (\ref{rnsb1}) and (\ref{rnsb2})  we have
\begin{equation}
 \lim_{t\to 1} R_{\rm N}t^{2}=-\frac{1}{8}
\end{equation}
indicating another dimensionless universal constant of $-1/8$.
Interestingly, it can be shown that this constant, obtained analytically, exactly agrees with the numerical result of the VdW fluid \cite{WeiWei}.

{\it Summary}
Starting from the Boltzmann entropy formula, we constructed the Ruppeiner geometry for a charged AdS black hole. Employing $T$ and $V$ as the fluctuation variables, we defined a new normalized scalar curvature of the geometry and showed that it provides useful information concerning black hole microstructure.  Previous work along these lines employed mass and pressure as the fluctuation variables \cite{WeiLiu}; in this case the corresponding scalar curvature does not diverge at the critical point, contrary to its use as a diagnostic of critical phenomena \cite{Ruppeiner}.

In comparison to a VdW fluid system, in which the weak attractive interaction dominates in all of parameter space above the coexistence curve, we found that a weak repulsive interaction dominates for charged AdS black holes at sufficiently high temperatures and small volumes (though elsewhere it is similar to a VdW fluid). Given previous work on the phase behavior of charged AdS black holes \cite{Kubiznak}, this surprising result indicates that important differences exist between this system and a VdW fluid at the microstructure level.

Furthermore, we obtained two other new universal properties of the microstructures for the charged AdS black hole. One is that the sign-changing curve demarcating the weak repulsive interactions (region III in Fig. \ref{pCAdStv}) is at fixed  $\tilde{V}$, independent of charge and temperature. Another is that the critical behavior of $R_{\rm N}$ indicates the existence of a universal critical exponent of $2$ along both the saturated small and large black hole curves, along with a universal coefficient of $-1/8$. We also have $R_{\rm N}\sim-\xi^{4}$, suggesting that other critical phenomena, such as critical opalescence, may have a geometric interpretation and more generally that the normalized scalar curvature $R_{\rm N}$ is indeed indicative of black hole microstructural properties.

In summary, novel information about black hole microstructures can be obtained from a geometric viewpoint. It would be interesting to apply our approach to other black hole systems with various properties, including angular momentum \cite{Altamirano:2014tva}, hair \cite{Giribet,Hennigar,Dykaar:2017mba}, and acceleration \cite{Anabalon:2018ydc,Anabalon:2018qfv}. The results will provide us with considerably more detailed information concerning the underlying degrees of freedom giving rise to black hole thermodynamics. Of further interest is an investigation of the physical laws and properties of the microstructures in the coexistence region, where the equation of state is not applicable.

{\emph{Acknowledgements}.}---This work was supported by the National Natural Science Foundation of China (Grants No. 11675064, No. 11875151, and No. 11522541) and the Natural Sciences and Engineering Research Council of Canada. S.-W. Wei was also supported by the Chinese Scholarship Council (CSC) Scholarship (201806185016) to visit the University of Waterloo.


\begin{thebibliography}{99}

\bibitem{Hawking}
 S. W. Hawking,
 {\em Particle creation by black holes},
   Commun. Math. Phys. \textbf{43}, 199 (1975).

\bibitem{Bekensteina}
 J. Bekenstein,
   {\em Black holes and the second law},
     Lett. Nuovo Cim. \textbf{4}, 737 (1972).

\bibitem{Bekensteinb}
 J. D. Bekenstein,
  {\em Black holes and entropy},
    Phys. Rev. D \textbf{7}, 2333 (1973).

\bibitem{Bardeen}
  J. M. Bardeen, B. Carter, and S. Hawking,
   {\em The four laws of black hole mechanics},
    Commun. Math. Phys. \textbf{31}, 161 (1973).

\bibitem{Vafa}
  A. Strominger and C. Vafa,
   {\em Microscopic origin of the Bekenstein-Hawking entropy},
     Phys. Lett. B \textbf{379}, 99 (1996).
    [arXiv:hep-th/9601029]

\bibitem{Maldacena}
 J. M. Maldacena and A. Strominger,
   {\em Statistical entropy of four-dimensional extremal black holes},
      Phys. Rev. Lett. \textbf{77}, 428 (1996),
      [arXiv:hep-th/9603060].

\bibitem{Callan}
 C. G. Callan and J. M. Maldacena,
  {\em D-brane approach to black hole quantum mechanics},
    Nucl. Phys. B \textbf{472}, 591 (1996),
    [arXiv:hep-th/9602043].

\bibitem{Horowitz}
   G. T. Horowitz and A. Strominger,
    {\em Counting states of near extremal black holes},
     Phys. Rev. Lett. \textbf{77}, 2368 (1996),
     [arXiv:hep-th/9602051].

\bibitem{Emparan}
   R. Emparan and G. T. Horowitz,
    {\em Microstates of a neutral black hole in M theory},
      Phys. Rev. Lett. \textbf{97}, 141601 (2006),
      [arXiv:hep-th/0607023]


\bibitem{Lunin}
   O. Lunin and S. D. Mathur,
    {\em AdS/CFT duality and the black hole information paradox},
      Nucl. Phys. B \textbf{623}, 342 (2002),
      [arXiv:hep-th/0109154].

\bibitem{Mathur}
   O. Lunin and S. D. Mathur,
    {\em Statistical interpretation of Bekenstein entropy for systems with a stretched horizon},
      Phys. Rev. Lett. \textbf{88}, 211303 (2002),
      [arXiv:hep-th/0202072].

\bibitem{Ruppeiner}
 G. Ruppeiner,
 {\em Riemannian geometry in thermodynamic fluctuation theory},
 Rev. Mod. Phys. \textbf{67}, 605 (1995); Erratum: Rev. Mod. Phys. \textbf{68}, 313 (1996).

\bibitem{Oshima}
 H. Oshima, T. Obata, and H. Hara,
  {\em Riemann scalar curvature of ideal quantum gases obeying Gentile's statistics},
   J. Phys. A: Math. Gen. \textbf{32}, 6373 (1999).

\bibitem{Kubiznak}
 D. Kubiznak and R. B. Mann,
  {\em $P$-$V$ criticality of charged AdS black holes},
     J. High Energy Phys. \textbf{1207}, 033 (2012), [arXiv:1205.0559[hep-th]].

\bibitem{Spallucci}
 E. Spallucci and A. Smailagic,
 {\em Maxwell's equal area law for charged Anti-de Sitter black holes},
 Phys. Lett. B \textbf{723}, 436 (2013), [arXiv:1305.3379[hep-th]].

\bibitem{Johnston}
  D. C. Johnston,
   {\em Thermodynamic properties of the van der Waals fluid},
  [arXiv:1402.1205[cond-mat.soft]].

\bibitem{WeiWei}
 S.-W. Wei, Y.-X. Liu, and R. B. Mann,
 {\em Ruppeiner geometry, phase transitions, and the microstructure of charged AdS black holes},
 in preparation.

\bibitem{WeiLiu}
 S.-W. Wei and Y.-X. Liu,
  {\em Insight into the microscopic structure of an AdS black hole from thermodynamical phase transition},
     Phys. Rev. Lett. \textbf{115}, 111302 (2015), [arXiv:1502.00386 [gr-qc]]; S.-W. Wei and Y.-X. Liu, Erratum: Phys. Rev. Lett. \textbf{116}, 169903(E) (2016).

\bibitem{Altamirano:2014tva}
  N. Altamirano, D. Kubiznak, R.~B. Mann and Z. Sherkatghanad,
{\em Thermodynamics of rotating black holes and black rings: phase transitions and thermodynamic volume},
  Galaxies {\bf 2}, 89 (2014), [arXiv:1401.2586 [hep-th]].

 \bibitem{Giribet}
G. Giribet, M. Leoni, J. Oliva, and S. Ray,
{\em Hairy black holes sourced by a conformally coupled scalar field in D dimensions},
Phys. Rev. D \textbf{89}, 085040 (2014), [arXiv:1401.4987 [hep-th]].

\bibitem{Hennigar}
R. A. Hennigar, E. Tjoa, and R. B. Mann,
{\em Thermodynamics of hairy black holes in Lovelock gravity},
J. High Energ. Phys. \textbf{1702}, 070 (2017), [arXiv:1612.06852 [hep-th]].

\bibitem{Dykaar:2017mba}
  H. Dykaar, R. A. Hennigar and R. B. Mann,
  {\em Hairy black holes in cubic quasi-topological gravity},
  JHEP {\bf 1705}, 045 (2017), [arXiv:1703.01633 [hep-th]].


\bibitem{Anabalon:2018ydc}
  A. Anabalon, M. Appels, R. Gregory, D. Kubiznak, R. B. Mann, and A. Ovgun,
{\em Holographic Thermodynamics of Accelerating Black Holes},
  Phys.\ Rev.\ D {\bf 98}, 104038 (2018), [arXiv:1805.02687 [hep-th]]


\bibitem{Anabalon:2018qfv}
   A. Anabalon, F. Gray, R. Gregory, D. Kubiznak, and R. B. Mann,
{\em Thermodynamics of Charged, Rotating, and Accelerating Black Holes},
  JHEP {\bf 1904}, 096 (2019), [arXiv:1811.04936 [hep-th]].

\end{thebibliography}
\end{document}